\documentclass[useAMS,  usenatbib]{mn2e}
\usepackage{tipa}
\usepackage{amsfonts}
\usepackage{amssymb}
\usepackage{graphicx}

\title[The stellar metallicity distribution in intermediate latitude fields
with BATC and SDSS data ]{The stellar metallicity distribution in
intermediate latitude fields with BATC and SDSS data}

\author[Xiyan Peng et al.]{Xiyan Peng$^{1}$,
 Cuihua Du$^{1}$
\thanks{E-mail:ducuihua@gucas.ac.cn},
Zhenyu Wu$^{2}$ \\
$^{1}$College of Physical Sciences, Graduate university of the
          Chinese Academy of Sciences, Beijing 100049, P. R. China\\
$^{2}$National Astronomical Observatories, Chinese
              Academy of Sciences, Beijing 100012, P. R. China}

\begin{document}

\date{Received}

\pagerange{\pageref{firstpage}--\pageref{lastpage}} \pubyear{2002}

\maketitle

\label{firstpage}

\begin{abstract}

Based on  the Beijing-Arizona-Taiwan-Connecticut (BATC)  and Sloan
Digital Sky Survey (SDSS) photometric data, we adopt SEDs fitting
method to \textbf{evaluate the metallicity distribution for
$\sim$40, 000 main-sequence stars in the Galaxy}. According to the
derived photometric metallicities  of these sample stars,  we find
that the metallicity distribution shift from metal-rich to
metal-poor with the increase of distance from the Galactic center.
\textbf{The mean metallicity is about of $-1.5\pm 0.2$ dex in the
outer halo and $-1.3\pm0.1$ dex in the inner halo. The mean
metallicity smoothly decreases from $-0.4$ to $-0.8$ in interval $0
< r \leqslant5 $ kpc. The fluctuation in the mean metallicity with
Galactic longitude can be found in interval $ 4 <  r \leqslant8 $
kpc. There is a vertical abundance gradients d[Fe/H]/dz $\sim -0.21
\pm 0.05$ dex kpc$^-$$^1$ for the thin disk ($z \leqslant 2 $ kpc).
At distance $2 < z \leqslant 5$ kpc, where the thick disk stars are
dominated, the gradients are about of $-0.16 \pm 0.06 $ dex
kpc$^-$$^1$, it can be interpreted as a mixture of stellar
population with different mean metallicities at all $z$ levels. The
vertical metallicity gradient is $ - 0.05 \pm 0.04$ dex kpc$^-$$^1$
for the halo ($z$ $>$ 5 kpc).} So there is little or no metallicity
gradient in the halo.

\end{abstract}

\begin{keywords}
Galaxy: structure-Galaxy: metallicity-Galaxy: formation.
 \end{keywords}

%___________________________________________________________________
%Section 1
\section{Introduction}

The structure, formation and evolution of the Galaxy are very
important issues in contemporary astrophysics. The basic components
of Galaxy are the thin disk, the thick disk, the halo and central
bulge, albeit that the inter-relationships and distinction among
different components remain subject to some debate (Gilmore 1983;
1984; Lemon et al. 2004). Recent studies, based on accurate
large-area surveys, have revealed that the Galaxy is marked by
numerous irregular substructure such as the Sagittarius dwarf tidal
stream in the halo (Ivezic et al. 2000; Yanny et al. 2000; Vivas et
al. 2001; Majewski et al. 2003) and the Monoceros stream closer to
the Galactic plane (Newberg et al. 2002; Rocha-Pinto et al. 2003).
Carollo et al. (2007) shown that the the halo is clearly divisible
into two broadly overlapping structural components - an inner and
outer halo from a local kinematic analysis. It is now apparent that
our Galaxy is much more complex system than we thought before.  The
formation of galaxies was long thought to be a steady process
resulting in a smooth distribution of stars (Bahcall \& Soneira
1981; Gilmore et al. 1989; Majewski 1993). But the view of the
formation of the Galaxy has changed dramatically since the
discoveries of complex substructures (Newberg et al. 2002; Belokurov
et al. 2007). The presence of these lumpy and complex substructure
are in qualitatively agreement with models for the formation of the
stellar halo through the accretion and merging of nearby dwarf
galaxies. Numerical simulations also suggest that this merger
process plays a crucial role in setting the structure and motions of
stars within galaxies (Bullock \& Johnston 2005).

The abundance distribution is particular importance to understanding
the formation and chemical evolution of the Galaxy (Freeman \&
Bland-Hawthorn 2002). Researchers have long sought to constrain
models for the Galactic formation and evolution on the basis of
observation of the stellar and clusters populations that it
contains. Specific models of galaxy formation make specific
predictions about the stellar abundance distribution. For example,
stars on more radial orbits are more metal-poor than stars on planar
orbits. This may indicate that the Milky Way formation began with a
relatively rapid collapse of the initial proto-galactic cloud, which
means the halo stars formed during the initial collapse,  the disk
stars formed after the gas had settled into the galactic plane.  But
the global collapse theory was unable to account for the lack of an
abundance gradient in the Galactic halo (Searle \& Zinn 1978). The
current view is that the Galactic halo formed at least partly
through the accretion of small satellite galaxies or merger of
larger systems (Freeman et al., 2002), which is well-supported by
studies of stellar kinematics and spatial distribution (Yannny et
al. 2003; Juric et al. 2008).  For the thick disk, it may be one of
the most significant components for studying signatures of galaxy
formation because it presents a snap frozen relic of the state of
the early disk (Freeman 2002). An intrinsic abundance gradient in
the thick disk would favor a scenarios which the thick disk was
formed either in the slow late stages of the early Galactic collapse
or the gradual kinematical diffusion of disk stars. On the contrary,
a irregular metallicity distribution or absence of gradient would
favor the thick disk having formed via the kinematical heating of
thin disk or from merger debris (Siegel et al. 2009).

The metallicity distribution of the Galaxy is best probed directly
through spectroscopic surveys (Yoss  et al. 1987; Allende-Prieto et
al. 2006). However, it has the advantage of using the photometric
metallicity of many more stars out to limiting magnitude of
photometric survey. Accurate determination of the properties of the
Galactic components requires surveys with sufficient sky coverage to
assess the overall geometry,   sufficient depth for mapping stars to
larger distance and sufficient information to obtain reasonable
distance estimates for these stars (De Jong et al. 2010). Over the
past few years,   numerous surveys have been used to investigate the
existence and size of the Galactic abundance gradient in the disk
and halo. The existence of a radial gradient in the Galaxy is now
well established. An average gradient of about $-0.06$ dex
kpc$^{-1}$ is observed in the Galactic disk for most of the elements
(Chen et al, 2003). De Jong et al. (2010) provided evidence for a
radical metallicity gradients in the Galactic stellar halo. However,
there is considerable disagreement about whether there is a vertical
metallicity gradient among field and/or open cluster stars of the
Galaxy.

The BATC multicolor photometric survey accumulated a large data base
which is very useful for studying the Galactic structure and
formation. Du et al. (2003) provided some information on the density
distribution of the main components of the Galaxy, which can present
constraints on the parameters of models of the Galactic structure.
Later, they use F and G dwarfs from the BATC data to study the
metal-abundance information (Du et al., 2004).  With the new
improved observation and improved knowledge regarding galaxy
formation, it becomes possible to further discuss the metallicity
gradient from different observation direction of the Galaxy.

In this paper, we attempt to study the metallicity gradient of the
Milky Way galaxy using the 21 BATC photometric survey fields
combined with the SDSS  photometric data. The outlines of this paper
is as follows: The BATC photometric system and data reduction are
introduced briefly in Section 2.  In Sect. 3 we describe the
theoretical model atmospheric spectra and synthetic photometry. The
metallicity distribution is discussed in Sect. 4. Finally in Sect. 5
we summarize our main conclusions in this study.

%___________________________________________________________________
%section 2
\section{Observations and data}
\subsection{BATC photometric system and SDSS photometric system}

The BATC survey performs photometric observations with a large field
multi-colour system.  There are 15 intermediate-band filters in the
BATC filter system, which covers an optical wavelength range from
3000 to 10000 {\AA} (Fan et al. 1996; Zhou et al. 2001).   The 60/90
cm f/3 Schmidt Telescope of National Astronomical Observatories
(NAOC) was used in the BATC program,   with a Ford Aerospace 2048
$\times$ 2048 CCD camera at its main focus. The field of view of the
CCD is $58^{\prime}$ $\times $ $ 58^{\prime}$ with a pixel scale of
$1\arcsec{\mbox{}\hspace{-0.15cm}.} 7$. The BATC magnitudes adopt
the monochromatic AB magnitudes as defined by Oke \& Gunn (1983).
The PIPELINE II reduction procedure was performed on each single CCD
frame to get the point spread function (PSF) magnitude of each point
source (Zhou et al. 2003). The detailed description of the BATC
photometric system and flux calibration of the standard stars can be
found in Fan et al. (1996) and Zhou et al. (2001, 2003). In order to
apply more color information to accurately estimate the photometric
stellar metallicity,  we combine the BATC colors with the SDSS
colors message for the sample stars.

The SDSS used a dedicated 2.5 m telescope which has an imaging
camera and a pair of spectrographs. The imaging camera (Gunn et al.
1998) contained 30 2048 $\times$ 2048 CCDs in the focal plane of the
telescope. The flux densities of observed objects were measured
almost simultaneously in five broad bands [ $u$,   $g$,   $r$, $i$,
$z$] (Fukugita et al. 1996; Gunn et al. 1998; Hongg et al. 2001).
For distinguishing explicitly between BATC and SDSS filter names, we
refer to the SDSS  filters and magnitudes as $u'$,   $g'$, $r'$,
$i'$, $z'$. The photometric pipeline (Luption et al. 2001) detected
the objects,   matched the data from the five filters and measured
instrumental fluxes,   positions and shape parameters. The shape
parameters allowed the classification of objects as  point source
 or extended. The magnitudes derived
from fitting a PSF are currently accurate to about 2 percents in
$g$, $r$ and  $i$, and 3-5 percents in $u$ and $z$ for bright ($\le$
20 mag) point source. In Table 1,    we list the parameters of the
BATC and SDSS filters. Col. (1) and Col. (2) represent the ID of the
BATC and SDSS filters,   Col. (3) and Col. (4) the central
wavelengths and FWHM of the 20 filters,   respectively. \textbf{The
reddening extinction for each star are determined from the SDSS
catalog. }

% Table 1_______________________________
\begin{table}
\begin{minipage}{120mm}
\caption{Parameters of the BATC and SDSS filters }
\begin{tabular}{cccc}\hline
\hline
No. & Filter &  Wavelength  & FWHM \\
    &         & (\AA)   & (\AA)   \\
\hline
1  & $a$ & 3371.5  & 359    \\
2  & $b$ & 3906.9  & 291    \\
3  & $c$ & 4193.5  & 309    \\
4  & $d$ & 4540.0  & 332    \\
5  & $e$ & 4925.0  & 374    \\
6  & $f$ & 5266.8  & 344    \\
7  & $g$ & 5789.9  & 289    \\
8  & $h$ & 6073.9  & 308   \\
9  & $i$ & 6655.9  & 491   \\
10 & $j$ & 7057.4  & 238   \\
11 & $k$ & 7546.3  & 192   \\
12 & $m$ & 8023.2  & 255   \\
13 & $n$ & 8484.3  & 167   \\
14 & $o$ & 9182.2  & 247   \\
15 & $p$ & 9738.5  & 275   \\
1  & $u'$ & 3543  & 569    \\
2  & $g'$ & 4770  & 1387    \\
3  & $r'$ & 6231  & 1373    \\
4  & $i'$ & 7625  & 1526    \\
5  & $z'$ & 9134  & 9500    \\
\hline \hline
\end{tabular}
\end{minipage}
\end{table}

\subsection{Direction and data reduction}
 Accurate determination of the
properties of components of the Milky Way  requires surveys with
sufficient sky coverage to assess the overall geometry (De Jong  et
al. 2010). Most previous investigations about the Galactic
metallicity distribution  use only  one or a few selected
lines-of-sight directions (Du et al. 2004;  Siegel et al. 2009;
Karatas et al. 2009). In this paper, the BATC photometry survey
presented 21 intermediate-latitude fields  in the multiple
directions.  These  fields used in this paper are towards the
Galactic center,   the anticentre,   the antirotation direction at
median and high latitudes, $|b|>35^{\circ}$.  Since metallicity
distribution at high Galactic latitudes are not strongly related to
the radial distribution, they are well suited to study the vertical
distribution of the Galaxy. Table 2 lists the locations of the
observed fields  and their general characteristics. In Table 2,
column 1 represents the BATC field name,  columns 2 and 3 represent
the Galactic longitude and latitude, columns 4 and 5 represent the
limit magnitude and  the number of sample star used in this study.
As shown in the Table 2,  the most photometric depth of our data is
21.0 mag in $i$ band. \textbf{In total, there are about 40, 000
sample stars in our study.}

 %  Table 2_______________________________
   \begin{table}

     \begin{minipage}{100mm}

     \caption{Relative information for the BATC
     observation fields}
         \begin{tabular}{ccccc}                 %{p{1cm}cp{1cm}cp{0cm}cp{0.5cm}p{1cm}c}
            \hline

            Observed field &  $l$ (deg) & $b$ (deg)  & $i$ (Comp) & star number  \\
  \hline

T485 &       175.7 &    37.8 &    21.0  & 1550 \\
T518 &       238.9 &    39.8 &    19.5  & 1584 \\
T288 &       189.0 &    37.5 &    20.0  & 2115 \\
T477 &       175.7 &    39.2 &    20.0  & 2001 \\
T328 &       160.3 &    41.9 &    19.5  & 1666 \\
T349 &       224.1 &    35.3 &    20.5  & 2436 \\
TA26 &       191.1 &    44.4 &    20.0  & 1285 \\
T291 &       167.8 &    46.4 &    20.0  & 1670 \\
T362 &       245.7 &    53.4 &    20.0  & 1237 \\
T330 &       147.2 &    68.3 &    20.5  & 1020 \\
U085 &       121.6 &    60.2 &    21.0  & 1679 \\
T521 &        56.1 &   -36.8 &    20.5  & 4430 \\
T491 &        62.9 &   -44.0 &    20.0  & 2824 \\
T359 &        79.7 &   -37.8 &    20.5  & 2932 \\
T350 &       251.3 &    67.3 &    19.5  & 1353 \\
T534 &        91.6 &    51.1 &    21.0  & 2044 \\
T193 &        59.8 &   -39.7 &    20.0  & 2830 \\
T516 &       125.0 &   -62.0 &    20.0  & 1113 \\
T329 &       169.9 &    50.4 &    21.0  & 1704 \\
TA01 &       135.7 &   -62.1 &    20.5  & 1342 \\
T517 &       188.6 &   -38.2 &    20.0  & 1296 \\

\hline
 \end{tabular}
\end{minipage}
\end{table}

 % Figure 1---------------------------------------

\begin{figure}
\begin{center}
\includegraphics[width=10cm]{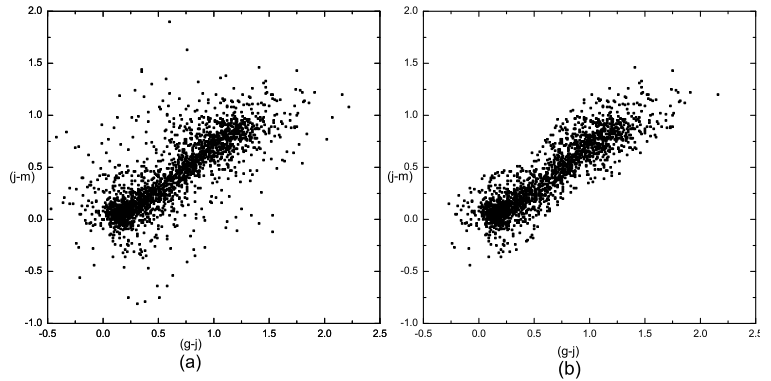}
\caption{Two-color diagrams for the BATC T329 field stars. The panel
(a) is  the distribution of sample stars in the ($j-m$)/($g-j$)
two-colour diagram, and  panel (b) is the ($j-m$)/($g-j$) two-colour
diagram after removing those objects that lay significantly off the
stellar locus. The line denotes the stellar locus from equation
(1).} \label{fig1}
\end{center}
\end{figure}

Here, because our fields in this work have also been observed by the
Sloan Digital Space Survey (SDSS-DR7) and each object type
(stars-galaxies-QSO) has been given. Thus, we can obtain a relative
reliable star catalogue.  In these star sample,   it is complete to
20.5 mag with an error of less than 0.1 mag in the BATC $i$ band.
Owing to the BATC observing strategy, data of stars brighter than
$m$ = 14 mag, are saturated, and star counts are not completed for
visual magnitudes fainter than $m$ = 21.0 mag. So our work is
restricted to the magnitudes range 15 $\leqslant $ $i$ $<$ 21. From
two-color diagrams for all objects in any field as an example, we
can find that most stars are plotted on a diagonal band. In Fig. 1
show the BATC ($g - j$) versus ($j - m$) two-color diagram for T329
field stars. The line which can be described by equation (1) donates
the stellar locus of the sample in the T329 field .

\begin{eqnarray}
(j - m) = -0.10245 + 0.81172 (g - j)
\end{eqnarray}

 The sample shows a main-sequence (MS) stellar
locus but has significant contamination from giants,   manifested in
the broader distribution overlying the narrow stellar locus. In
order to determine the MS star sample,   we use multi-color
selection criteria outlined in Karaali et al. (2003) and Juric et
al. (2008), which remove objects based on their location relative to
the dominant stellar locus. For example,    Juric et al. (2008)
applied an extra procedure which consists of rejecting objects at
distance larger than 0.3 mag from the stellar locus in order to
remove hot dwarfs, low-red shift quasars,   and white /red dwarf
unresolved binaries from their sample. The procedure does well for
high latitude field data from SDSS (Karaali et al. 2003;  Yaz et al.
2010). Fig. 1b gives the cleaning sample stars after rejecting those
objects which is lay significantly off the stellar locus.

\section{THEORETICAL MODEL AND CALIBRATION FOR METALLICITY}
\label{sect:data}
\subsection{ theoretical stellar library and synthetic photometric}
A homogeneous and complete stellar library can match any ambitious
goals imposed on a standard library.  Lejeune et al. (1997) % reference %
 presented a hybrid library of synthetic stellar spectra. The library covers a
wide range of stellar parameters $T_{\rm eff}$= 50, 000 K to 2, 000
K in intervals of 250 K, log~$g$= $-1.02$ to 5.50 in main increments
of 0.5, and [M/H]= $-5.0$ to +1.0. For each model in the library, a
flux spectrum is given for the same set of 1221 wavelength points
covering the range 9.1 to 160, 000 nm, with a mean resolution of
20{\AA} in the visible. The spectra are thus in a format which has
proved to be adequate for synthetic photometry of wide- and
intermediate-band systems (Du et al. 2004).

On the basis of the theoretical library, we calculate synthetic
colors of the BATC  and SDSS photometric system. Here, we synthesize
colors for simulated stellar spectra with $T_{\rm eff}$ and log~$g$
characteristic of main sequences (log~$g$ = 3.5, 4.0, 4.5 for
dwarfs) and 19 values of metallicity ([M/H]=  $-5.0, -4.5, -4.0,
-3.5$, $-3.0, -2.5, -2.0, -1.5, -1.0, -0.5$,$-0.3, -0.2, -0.1$, 0.0,
+0.1, +0.2, +0.3, +0.5 and +1.0), where [M/H] denotes metallicity
relative to hydrogen. The synthetic $i$th filter magnitude can be
calculated with equation (2).
\begin{equation}
 m = -2.5~{\rm log}\frac{\int{F_{\lambda}\phi_{i}({\lambda}){\rm d}\lambda}}
{\int{\phi_{i}({\lambda}){\rm d}\lambda}} - 48.60,
\end{equation}
Where ${F_{\lambda}}$ is the flux per unit wavelength, $\phi_{i}$ is
the transmission curve of the $i$th filter of the BATC  or SDSS
filter system (Du et al. 2004) .

The bluer colors are sensitive to metallicity down to the lowest
observed metallicities because most of the line-blanketing from
heavy elements occurs in the shorter wavelength regions. In
contrast, the redder colors are primarily sensitive to temperature
index. The BATC $a$, $b$ bands contain the Balmier jump, a stellar
spectral feature which is sensitive to surface gravity. Since our
sample includes only main sequences, it conveys little gravity
information.  It should be mentioned that, although the metallicity
or temperature derived from synthetic photometry is not very
accurate for a single star, perhaps which can be distorted by a poor
point, it is meaningful for the statistic analysis of sample stars.

\subsection{ METALLICITY AND PHOTOMETRIC PARALLAX }
The most accurate measurements of stellar metallicity are based on
spectroscopic observation. Despite the recent progress in the
availability of stellar spectra (e.g. SDSS-III and RAVE), the
stellar number detected in photometric surveys is much more than
spectroscopic observation. So photometric methods have also often
been used to give the stellar metallicity.  For example,    Sandage
(1969) detailed a technique using UBV photometry indices to measure
approximate abundance. Karaali (2003) evaluated the metal abundance
by ultraviolet-excess photometric parameter using CCD UBVI data.
Karaali (2005) extended this method to the SDSS photometry. Ivezic
et al. (2008) obtained the mean metallicity of stars as a function
of $u-g$ and $g-r$ colors of SDSS data. For the BATC multicolor
photometric system,   there are 15 intermediate-band filters
covering an optical wavelength range from 3000 to 10000 \AA. There
are 5 filters for the SDSS photometric system. So the SEDs of 20
filters for every object are equivalent to a low resolution
spectrum.

The sample SEDs simulation with template SEDs can be used to derive
the parameter of sample stars (Du et al. 2004). The standard
minimization,   computing and minimizing the deviations between the
photometric SEDs of the star and the templates SEDs obtained with
the same photometric system,   is used in the fitting process. The
minimum indicated the best fit to the observed SED by the set of
spectra (Du et al. 2004).

\begin{equation}
{\chi^{2}=\sum\limits_{l=1}^{N_{filt}=20}\left [\frac{m_{obs,l}-
 m_{temp,l}} {\sigma_{l}} \right ]^{2}},
\end{equation}
where ${m_{obs,l}}$, $m_{temp,l}$ and $\sigma_{l}$ are the observed
magnitude, template magnitude and their uncertainty in filter $l$,
respectively, and $N_{filt}$ is the total number of filters in the
photometry.

\textbf{ According to the results of SEDs fitting,   the metallicity
and temperature of about 40, 000 sample stars are obtained in 21
fields. In addition, we extract the spectroscopic metallicities from
sdss DR7 database for our studied 21 fields, and there are about 870
stars for which they also have photometric metallicities from our
method. Using these stars, we present a calibration of our SED
fitting method. After applying calibration, it is reliable for the
derived photometric metallicities from SEDs fitting method. In
Figure 2., we present the difference of photometric and
spectroscopic metallicity as a function of $(g - r)$ color. } The
uncertainties of metallicity obtained from comparing SEDs between
photometry and theoretical models are due to the observational error
and the finite grid of the models. For the metal-poor stars
([Fe/H]$<-1.0$), the metallicity uncertainty is about 0.5 dex, and
0.2 dex for the stars [Fe/H]$>-0.5$ (Du et al. 2004). The
metallicity distribution diagram for all sample stars was given in
Fig. 3. One local maximum appear at [Fe/H] from -0.5 to 0 dex,   and
a tail down to \textbf{-3.0 dex}.

\begin{figure}
\begin{center}
\includegraphics[width=9cm]{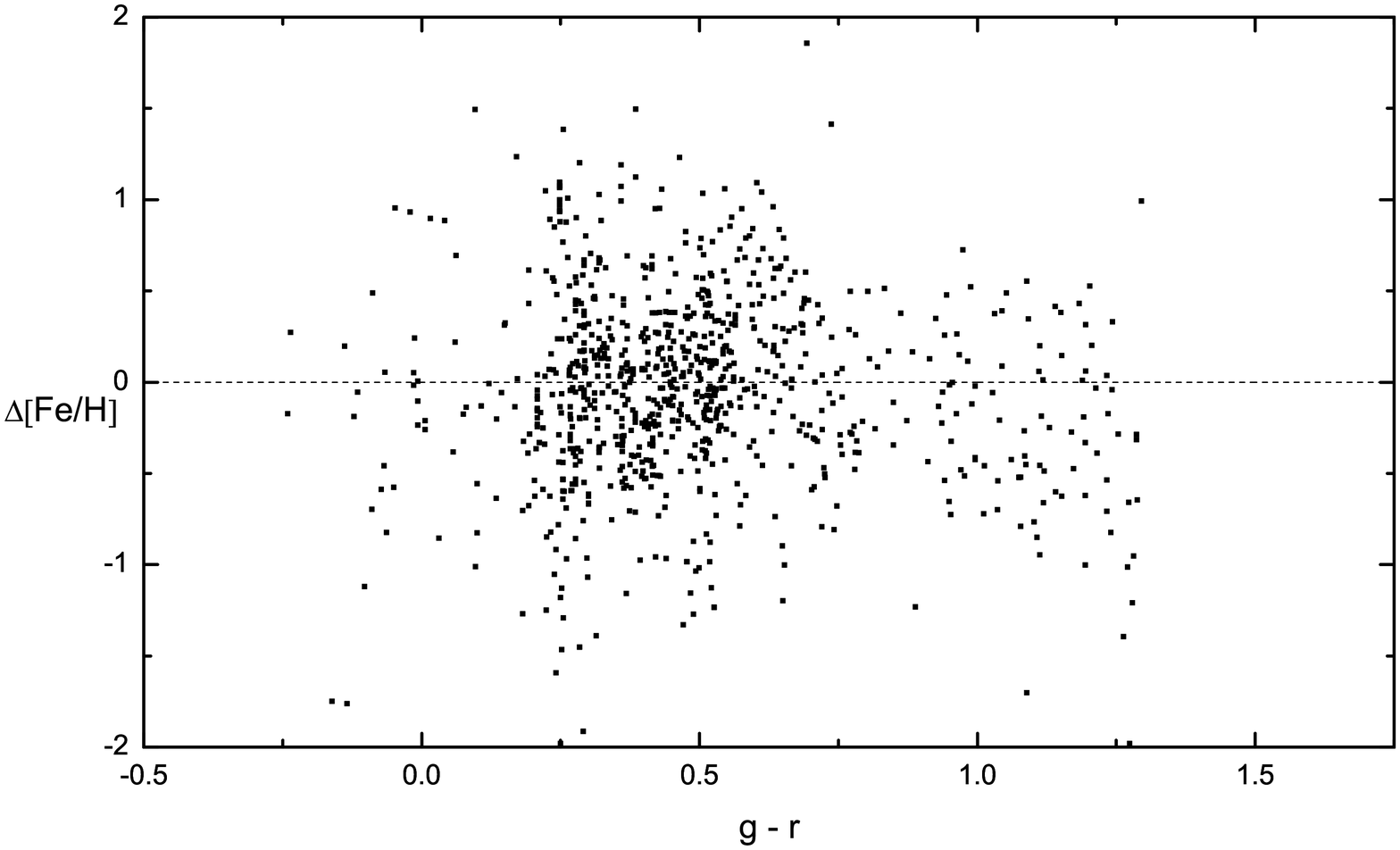}
\caption{The difference of photometric and spectroscopic metallicity
for 870 stars as a function of $(g - r)$ color is shown.}
\label{fig2}
\end{center}
\end{figure}

\begin{figure}
\begin{center}
\includegraphics[width=9cm]{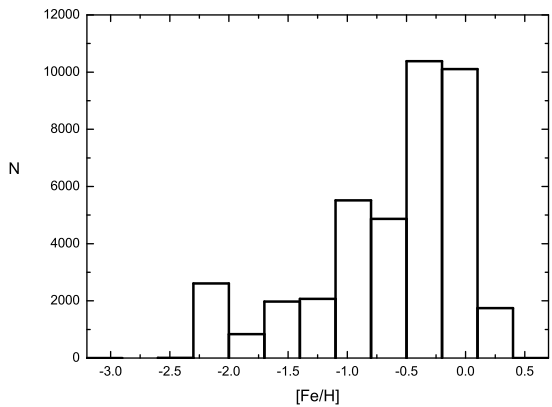}
\caption{The metallicity distribution for all the sample stars
selected in our study is shown. } \label{fig3}
\end{center}
\end{figure}

The stellar type can be derived from effective temperature of
dwarfs, then the stellar distance relative to the sun can be
obtained by equation (4).
\begin{eqnarray}
m_{v}-M_{v}=5lgr - 5 + A_{v}
\end{eqnarray}
where $m_{v}$ is  the visual magnitude,   absolute magnitude $M_{v}$
can be obtained according to the stellar type. The reddening
extinction $A_{v}$ is small for most fields. We adopted the absolute
magnitude versus stellar type relation for main-sequence stars from
Lang (1992).  $r$ is the stellar distances. The vertical distance of
the star to the galactic plane can be evaluated by equation (5):
\begin{eqnarray}
z = r sin b
\end{eqnarray}

A variety of errors affect the determination of stellar distances.
The first source of errors is from photometric uncertainty less than
0.1 mag in the BATC $i$ band; the second from the misclassification,
which should be small due to the multicolor photometry. For
luminosity class V, types F/G, the absolute magnitude uncertainty is
about 0.3 mag. In addition, there may exist an error from the
contamination of binary stars in our sample. We neglect the effect
of binary contamination on distance derivation due to the unknown
but small influence from mass dsitribution in binary components
(Kroupa et al.~1993; Ojha et al.~1996).

\begin{figure*}
\includegraphics[width=18cm]{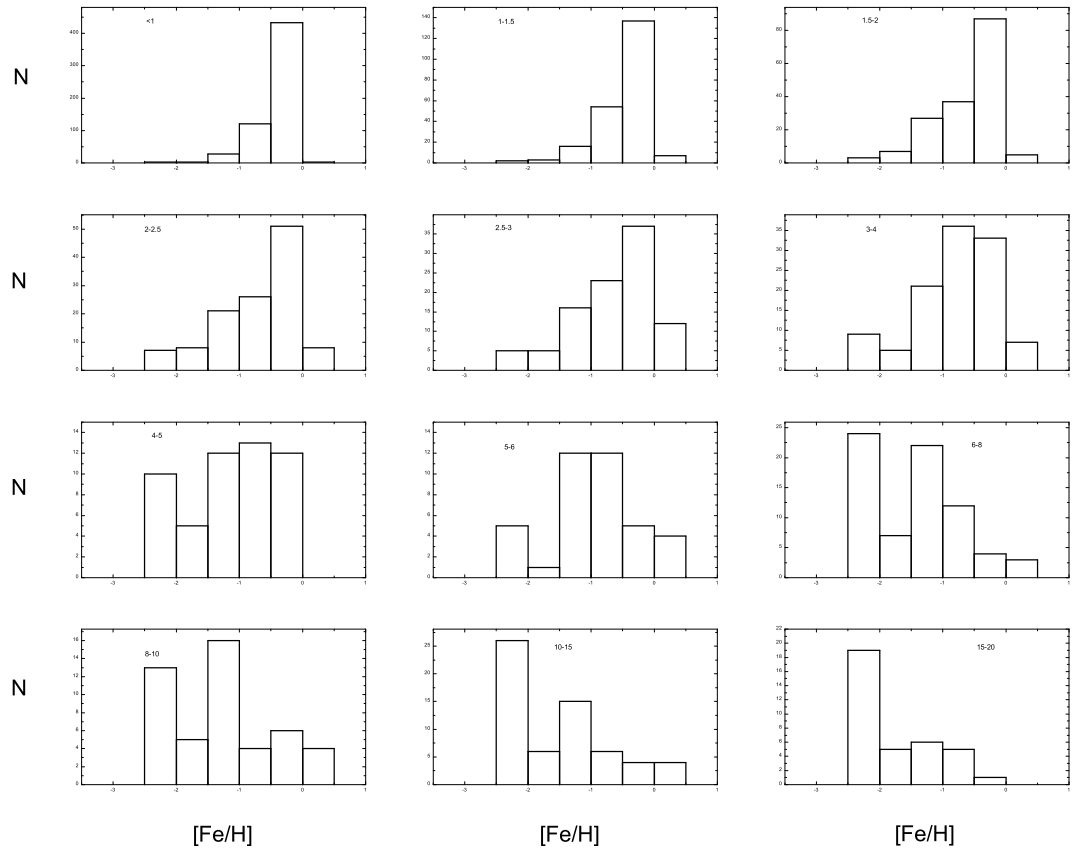}
\caption{ The metallicity distribution for the T291 field in
different distance range is shown. In the short distance,  r $<$
1kpc, most stars are in the range [0,  $-0.5$],   in the larger
distance, 15 $<$ r $<$ 20 kpc, most stars are poorer than $-1$. }
\label{fig3}
\end{figure*}

\section{METALLICITY DISTRIBUTION }

It is well known that the chemical abundance of stellar population
contains much information about the population$^,$s early evolution.
The stellar metallicity distribution in the Galaxy has been the
subject of photometric and spectroscopic surveys (Gilmore et
al.~1985; Ratnatunga et al.~1989; Friel~1988). In this study, we
want to explore  possible stellar metallicity distribution variation
with the observation direction.  A method of SED combination for the
SDSS and BATC photometries has been adopted to give the steller
metallicity distribution.   At first, the metallicity for the sample
stars can be derived by comparing SEDs between photometry and the
theoretical models. The SED fitting method is described in Section
3. 2.  The mean metallicity distribution for each field is
determined in the following distance intervals (in kpc): $15 <
r\leqslant 20$, $10< r \leqslant 15$, $8<  r\leqslant 10$, $6<  r
\leqslant 8$, $5< r\leqslant 6$, $4< r\leqslant 5$, $3< r\leqslant
4$,
  $2.5<  r\leqslant 3$,    $2<  r \leqslant2.5$,
  $1.5<  r\leqslant 2$,    $1<  r \leqslant1.5$,
  $0.25<  r\leqslant 1$.
As an example the metallicity distributions as a function of
vertical distance for the field T291 is presented in Fig. 4.   From
the figures (Fig. \,4), it is clear  that there is a number-shift
from metal-rich stars to metal-poor ones with the increasing of
distance. In star counts the younger metal rich stars are confined
to regions close to the Galactic mid-plane, while the older,
metal-poorer stars with a larger scale height dominated at larger
vertical distances from the Galactic plane.

\subsection{Metallicity variation with Galactic longitude}
Mean metallicity distribution as a function of Galactic longitude
for different distance intervals  are presented in Fig. 5.  The
 mean metallicity shift from metal-rich to metal-poor with the increase of
distance from the Galactic center can be found in Fig. 5.  The solid
points represent the south galactic latitude fields, The open square
points represent the  north galactic latitude fields. \textbf{As
shown in Fig. 5, the mean metallicity in interval $10 < r \leqslant
20$ kpc is around of $-1.5$ dex. In intervals $8 <   r \leqslant10 $
kpc our result indicates that the mean metallicity is  $\sim$
$-1.3$. The mean metallicity in intervals $5 <  r \leqslant 8$ kpc
is about $-1.0$ and the mean metallicity smoothly decreases from
$-0.6$ to $-0.8$ in intervals $2.5 < r \leqslant5 $ kpc. Our results
are consistent with the results of Siegel et al. (2009) and Karatas
et al. (2009). Siegel et al. find a monometallic thick disk and halo
with abundances of [Fe/H] = $-0.8$  and $-1.4$ respectively. Karatas
et al. derive mean abundance values of [Fe/H]= $-0.77 \pm 0.36$ dex
for the thick disk, and [Fe/H] = $-1.42 \pm 0.98$.  The mean
metallicity decreases from $-0.4$ to $-0.6$ in intervals $0 < r
\leqslant 2.5 $ kpc. The mean metallicity in interval $0.25 < r
\leqslant 1 $ kpc is [Fe/H] $\sim$ $-0.3$, which is consistent with
the result of Yaz et al. (2010).}

As shown in Fig. 5,  at larger distance, r $>$ 10 kpc, compared to
the typical error bars, the mean metallicity distributions variation
with Galactic longitude is almost flat.\textbf{ For $ 4 <  r
\leqslant8 $ kpc, there is a fluctuation  in the mean metallicity
with Galactic longitude. The overall distribution of mean
metallicity has a maxminum at $l$ $\sim$ 200$^\circ$. For $ 2 <  r
\leqslant8 $ kpc, the T517 filed (Galactic coordinates :
$l=188.6^{\circ}, b=-38.2^{\circ}$; Equatorial coordinates : $\alpha
= 58.59^{\circ}$, $\delta = -0.35^{\circ}$) and the TA26 field
(Galactic coordinates : $l=191.1^{\circ}, b=44.4^{\circ}$;
Equatorial coordinates : $\alpha = 139.956^{\circ}$, $\delta =
33.745^{\circ}$) show metal rich character related to other fields.
This feature may reflect a fluctuation from streams (such as
Monocers stream) which are accreted from nearby galaxies. Juric et
al.(2008) detect two overdensities in the thick disk region. Klement
et al.(2009) also find individual stream from the SSPP in the
direction with central coordinates (Equatorial coordinates):
$\alpha$ = $58.58^{\circ}$ and $\delta$ = $-4.99^{\circ}$. Maybe the
deviant behaviors of the two fields result  from systematical error
in the observation.} In the work of AK et al. (2007), they find that
the metallicity distributions for both (relatively) short and large
vertical distances show systematic fluctuations. The scaleheight of
thick disk varies with the observed direction were found in the
works of Du et al. (2006) and Bilir et al. (2008).
\begin{figure*}

\includegraphics[width=15cm]{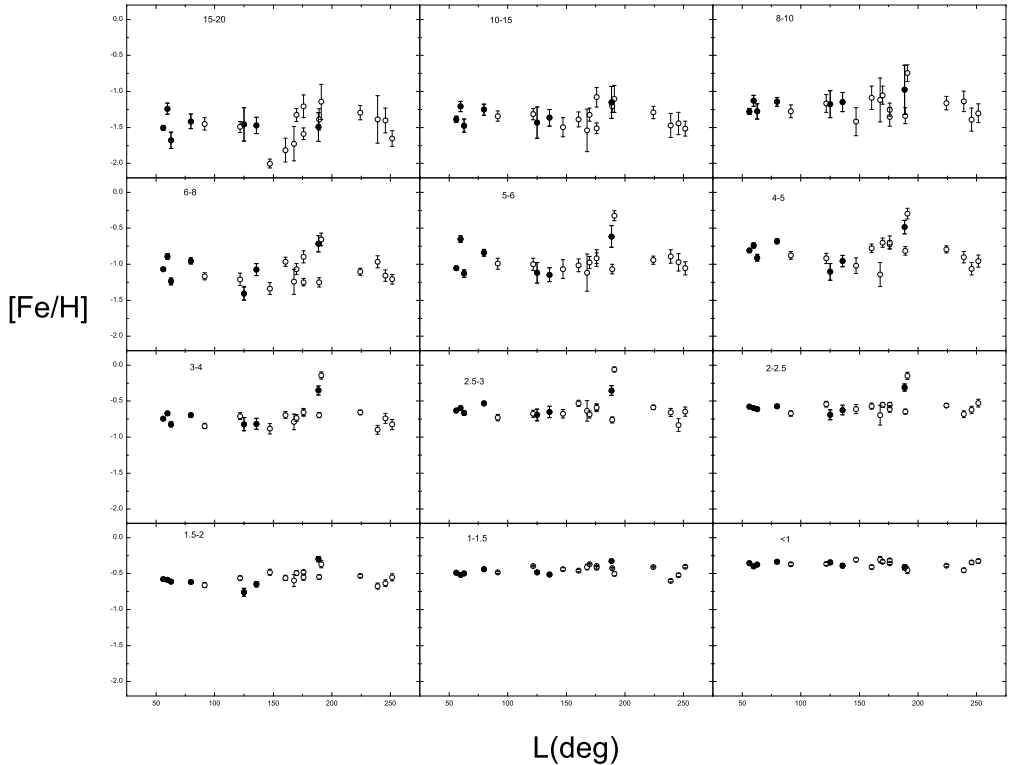}
\caption{The mean metallicity distribution as a function of galactic
longitude in different distance intervals are shown.  }

\end{figure*}

\subsection{THE VERTICAL METALLICITY GRADIENT }
 Detailed information about the vertical metallicity
gradient can provide important clue about the formation scenario of
stellar population.  Here, we used the mean metallicity to described
the metallicity distribution function. As an example,   The
distribution trend of mean metallicity [Fe/H] with  height above the
galactic plane [$z$] for the T291 field is shown in Fig. 6. The
metallicity gradients for all the fields in different $z$ intervals
$z < 2$ kpc, $2  < z \leqslant 5$ kpc and $5  < z \leqslant 15$ kpc
are given  in Fig. 7 and detailed in Table 3. In Table 3, Column 1
represents the BATC field name, Columns 2 - 7 represent gradient and
error of gradient in different $z$ distance: $z \leqslant 2$ kpc, $2
< z \leqslant 5$ kpc and $5  < z \leqslant 15$ kpc, respectively.

From the Fig. 7 we can find that the variation of the  gradient for
the halo with galactic longitude is flat  and the mean gradient of
halo is about $-0.05 \pm 0.04$ dex kpc$^-$$^1$ ($5 < z \leqslant 15$
kpc), which is essentially in agreement with the conclusion of Yaz
et al. (2010)  and Du et al. (2004).  Du et al.(2004) find the small
or zero gradient d[Fe/H]/dz = $-0.06 \pm 0.09$ in the halo. Yaz et
al. (2010) find $d[M/H]/dz = -0.01$ dex kpc$^-$$^1$ for the inner
spheroid. The result of Karaali et al. (2003) is slightly steeper
than the value of  our result.  Karaali et al. (2003) find that
there is a metallicity gradient d[Fe/H]/dz $\sim -0.1$ dex
kpc$^-$$^1$ in the inner halo ($5 < z \leqslant 8$ kpc ) and zero in
the outer part ($8 < z \leqslant 10$ kpc).     From Fig. 6 we can
find that the incompleteness of the star sample causes significant
statistical uncertainties at large distance. Probably, there is
little or no metallicity gradient in the halo. It is consistent with
the merger or accretion origin of the outer halo.

As shown in Fig. 7, at distance $0 < z < 2 $ kpc, the mean vertical
abundance gradient is about  d[Fe/H]/dz $\sim -0.21 \pm 0.05$ dex
kpc$^-$$^1$. The  value for the vertical metallicity at distance $0
< z < 2$ kpc is in agreement with the canonical metallicity
gradients with the same $z$ distances. For example,  Yaz et al.
(2009) find the metallicity gradient is d[Fe/H]/dz $\sim -0.3$ dex
kpc$^-$$^1$ for short distance. The metallicity gradient is found to
be d[Fe/H]/dz $\sim -0.37$ dex kpc$^-$$^1$ for $z$ $<$ 4 kpc in the
work  of Du et al. (2004). The result of Karaali et al. (2003) can
be described as d[Fe/H]/dz $\sim -0.2$ dex kpc$^-$$^1$ for the thin
and thick disk.

At distance $2 < z \leqslant5$ kpc, where the thick disk stars
dominated, the gradient is about $-0.16 \pm 0.06 $ dex kpc$^-$$^1$
in our work which is consistent with the work of karaali et al.
(2003) and less than the value of Du et al. (2004). Du et al. (2004)
point out that the metallicity gradient is d[Fe/H]/dz $\sim$ $-0.37$
dex kpc$^-$$^1$. In our study, the thick disk gradient is
interpreted as different contribution from three components of the
Galaxy at different $z$ distance. The existence of a clear vertical
metallicity of the thick disk would be an important clue about the
origin of the thick disk. However, it is an open question for the
formation of the thick disk component. A number of models have been
put forward since the confirmation if its existence.  Chen et al.
(2001) support that the thick disk formed through the heating of a
preexisting thin disk, with the heating mechanism being the merging
of a satellite galaxy. Here, we also favor the thick disk having
formed via the kinematical heating of thin disk and from merger
debris. Thus, there is a irregular metallicity distribution or
absence of intrinsic gradient.

\begin{figure}
\begin{center}
\includegraphics[width=9cm]{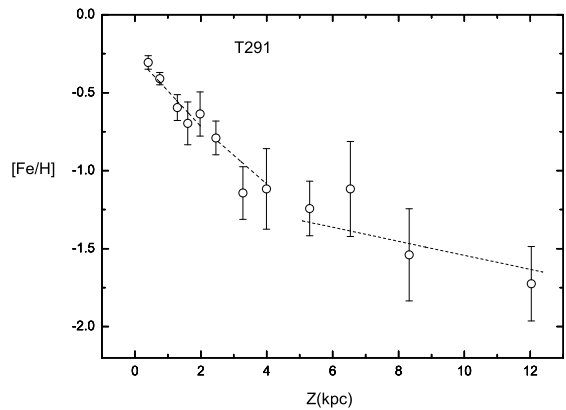}
\caption{The mean  metallicity as a function of vertical distance
$z$ for the T291 field. The metallicity gradients of the thin disk,
thick disk and halo are  $ -0.23\pm0.03$,   $ -0.18\pm 0.07$,
$-0.05\pm0.01$ dex kpc$^-$$^1$,  respectively. } \label{fig1}
\end{center}
\end{figure}

\begin{figure*}

\includegraphics[width=15cm]{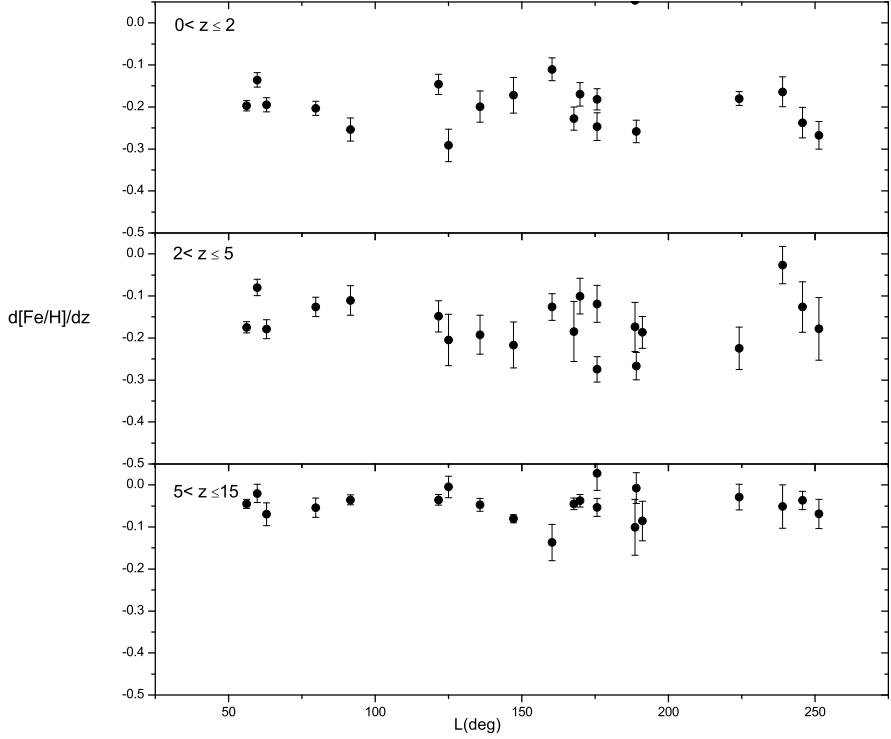}
\caption{The  metallicity gradients distribution for all fields in
this study are shown for the intervals $z$ $<$ 2 kpc , 2 kpc $<$ $z$
$<$ 5kpc and 5 kpc $<$ $z$ $<$ 15kpc.} \label{fig1}

\end{figure*}

%  Table 2_______________________________
   \begin{table*}
    \centering
     \begin{minipage}{100mm}
     \caption{ The gradient distribution in different distance interval for the selected fields}
         \begin{tabular}{cccccccc}
          \hline
             & 0-2 (kpc) &  &2-5 (kpc)&  & 5-15 (kpc)&   \\
           \hline
      Observed field & gradient & error &gradient& error & gradient & error  \\
 \hline

 T193 &       -0.136 &      0.017 &    -0.080 &      0.020 &    -0.020 &   0.022 \\
 T288 &       -0.258 &      0.027 &    -0.267 &      0.032 &    -0.007 &   0.037 \\
 T291 &       -0.228 &      0.028 &    -0.185 &      0.071 &    -0.045 &   0.014 \\
 T328 &       -0.110 &      0.027 &    -0.126 &      0.032 &    -0.137 &   0.043 \\
 T329 &       -0.170 &      0.028 &    -0.100 &      0.042 &    -0.037 &   0.015 \\
 T330 &       -0.171 &      0.042 &    -0.217 &      0.055 &    -0.080 &   0.010 \\
 T349 &       -0.180 &      0.017 &    -0.225 &      0.050 &    -0.029 &   0.031 \\
 T350 &       -0.267 &      0.033 &    -0.178 &      0.074 &    -0.069 &   0.035 \\
 T359 &       -0.203 &      0.017 &    -0.126 &      0.023 &    -0.054 &   0.023 \\
 T362 &       -0.237 &      0.037 &    -0.126 &      0.060 &    -0.037 &   0.022 \\
 T477 &       -0.181 &      0.025 &    -0.275 &      0.030 &    -0.053 &   0.022 \\
 T485 &       -0.247 &      0.033 &    -0.119 &      0.044 &     0.028 &   0.040 \\
 T491 &       -0.194 &      0.017 &    -0.179 &      0.020 &    -0.070 &   0.027 \\
 T516 &       -0.291 &      0.039 &    -0.204 &      0.061 &    -0.005 &   0.026 \\
 T517 &        0.054 &      0.041 &    -0.173 &      0.059 &    -0.101 &   0.067 \\
 T518 &       -0.163 &      0.035 &    -0.026 &      0.045 &    -0.051 &   0.052 \\
 T521 &       -0.197 &      0.012 &    -0.175 &      0.013 &    -0.045 &   0.011 \\
 T534 &       -0.254 &      0.028 &    -0.110 &      0.035 &    -0.035 &   0.012 \\
 TA01 &       -0.199 &      0.037 &    -0.192 &      0.046 &    -0.047 &   0.015 \\
 TA26 &        0.307 &      0.030 &    -0.187 &      0.038 &    -0.085 &   0.047 \\
 U085 &       -0.146 &      0.024 &    -0.148 &      0.037 &    -0.035 &   0.013 \\

\hline
 \end{tabular}
\end{minipage}
\end{table*}

\section{CONCLUSIONS AND SUMMARY}

In this work, based on the BATC and SDSS photometric data, we
evaluated the stellar metallicity  distribution for  40, 000
main-sequence stars in the Galaxy by adopting SEDs fitting method.
These selected fields  are towards the Galactic center, the
anticentre, the antirotation direction at median  and high latitudes
.  The metallicity distribution could be obtained up to distances $r
= 20$ kpc,   which covers the thin disk, thick disk and halo. We
determined the mean stellar metallicity as a function of vertical
distance in different direction. It can be clearly seen that the
metallicity distribution shift from metal-rich to metal-poor with
the increase of distance from the Galactic center. The mean
metallicity is about $-1.5\pm 0.2$ dex in intervals $10 < r
\leqslant 20$ kpc and $-1.3\pm0.1$ dex in interval $8 < r
\leqslant10 $ kpc. The mean metallicity smoothly decreases from
$-0.6$ to $-0.8$ in interval $2.5< r \leqslant5 $ kpc, while the
mean metallicity decreases from $-0.4$ to $-0.6$ in interval $0 < r
\leqslant 2.5 $ kpc. In addition, a fluctuation in the mean
metallicity with Galactic longitude can be found and the overall
distribution has a maximum at about $l$ $\sim$ 200$^\circ$ in
interval $4 < r \leqslant8 $ kpc. Maybe this feature can be related
with the substructure or streams (such as Monoceros stream) which
are accreted from nearby galaxies.  At the same time, we find the
vertical abundance gradients for the thin disk ($0 < z < 2 $ kpc) is
d[Fe/H]/dz $\sim -0.21 \pm 0.05$ dex kpc$^-$$^1$, and a vertical
gradient $-0.16 \pm 0.06 $ dex kpc$^-$$^1$  at distances $2<  z
\leqslant 5$ kpc where the thick disk stars are dominated. Here, we
consider the thick disk gradient may be the result from the
different contributions from three components of the Galaxy at
different $z$ distance. The vertical gradient d[Fe/H]/dz $\sim -
0.05 \pm 0.04$ dex kpc$^-$$^1$  is found in distance $5 < z
\leqslant 15$ kpc. So, there is little or no gradient in the halo.
These results are in agreement with the values in the literature
(Yoss et al.~1987; Trefzger et al.~1995; Karaali et al. 2003; AK et
al. 2007; Yaz et al. 2010). It is possible that additional
observational investigations (some projects aimed at spectroscopic
sky surveys such as SEGUE, LAMOST, GAIA) will give more evidence for
the metallicity gradient of the Galaxy and therefore provide a
powerful clue to the disk and halo formation.

\section*{Acknowledgments}
We especially thank the anonymous referee for numerous helpful
comments and suggestions which have significantly improved this
manuscript. This work was supported by the joint fund of Astronomy
of the National Nature Science Foundation of China and the Chinese
Academy of Science, under Grants 10778720 and 10803007. This work
was also supported by the GUCAS president fund.

\label{lastpage}

\end{document}